# HIGH-MEMORY MASKED CONVOLUTIONAL CODES FOR POST-QUANTUM CRYPTOGRAPHY

Meir Ariel

School of Electrical and Computer Engineering
Tel Aviv University, Israel

### *ABSTRACT*

*This paper presents a novel post-quantum cryptosystem based on high-memory masked convolutional codes. Unlike conventional code-based schemes that rely on block codes with fixed dimensions and limited error-correction capability, our construction offers both stronger cryptographic security and greater flexibility. It supports arbitrary plaintext lengths with linear-time decryption and uniform per-bit computational cost, enabling seamless scalability to long messages. Security is reinforced through a higher-rate injection of random errors than in block-code approaches, along with additional noise introduced via polynomial division, which substantially obfuscates the underlying code structure. Semi-invertible transformations generate dense, random-like generator matrices that conceal algebraic properties and resist known structural attacks. Consequently, the scheme achieves cryptanalytic security margins exceeding those of the classic McEliece system by factors greater than $2^{100}$. Finally, decryption at the recipient employs an array of parallel Viterbi decoders, enabling efficient hardware and software implementation and positioning the scheme as a strong candidate for deployment in practical quantum-resistant public-key cryptosystems.*

### *KEYWORDS*

*Code-based cryptography, Post-quantum cryptography, Convolutional codes*

## 1. INTRODUCTION

Code-based cryptography, introduced by Robert McEliece [1] in 1978, employs binary Goppa codes to build a public-key encryption scheme. The core concept disguises an error-correcting code using invertible linear transformations. The sender deliberately introduces errors into the ciphertext (i.e., the codeword), making decryption challenging for attackers while allowing the recipient, who possesses the private key, to correct the errors and retrieve the plaintext.

Since its inception, numerous modifications to the McEliece cryptosystem have emerged, primarily by employing alternative code families. However, most of these failed to preserve security at a level comparable to the original scheme [2-3]. As a result, Classic McEliece advanced to Round 4 of the NIST post-quantum cryptography (PQC) standardization process [4], but was not ultimately selected for final adoption. Comprehensive surveys of code-based cryptosystems and their role in the NIST PQC process are provided in [5–6]. Despite its solid theoretical foundation, Classic McEliece suffers from practical limitations, most notably its reliance on fixed Goppa codes with relatively rigid error-correction capabilities—essentially limited to parameter sets such as $(N, K, t) = (1024, 524, 50)$ and $(4096, 3556, 45)$, where $K \times N$ denotes the dimensions of the generator matrix and $t$ the error correction capacity. These constraints restrict adaptability to different security levels and hinder broader deployment.

Our work introduces a novel PQC approach using high-memory masked convolutional codes, offering several advantages over traditional code-based cryptosystems:

**Diverse Code Selection**: A wide range of convolutional codes can form public and private keys, allowing customization to meet specific performance and security needs.

**Stronger Public Key Security**: Our method's high-density generating matrix with a random-like structure significantly improves security against cryptanalysis compared to the low-density matrices of other code-based systems [7]. The flexibility to select convolutional codes with strong error-correcting capabilities enables the introduction of a higher (but unknown) number of random errors, boosting cryptanalysis resistance by factors exceeding $2^{100}$, depending on key length, compared to Classic McEliece.

**Scalable Key Length and Decoding Complexity**: Unlike block codes with fixed dimensions, our scheme supports plaintexts of arbitrary length, meeting diverse security needs. The recipient's decoding complexity scales linearly with key length.

**Efficient Software/Hardware Implementation**: The decoding process utilizes an array of Viterbi decoders, based on directed graph methods, facilitating efficient and straightforward hardware and software implementation.

The paper is structured as follows: Section 2 describes the construction of public and private keys, forming our cryptographic framework's core. Section 3 covers the sender's encryption process, while Section 4 details the recipient's decryption method. Section 5 examines potential eavesdropping attacks and their impact. Section 6 discusses the rationale behind our polynomial selection strategy and the expected error rate, crucial for security. Section 7 provides a worked example demonstrating practical implementation. Finally, Section 8 analyses resistance to cryptanalysis and complexity metrics, highlighting security benefits over traditional block-code-based systems, and Section 9 concludes with key findings.

## 2. PUBLIC AND PRIVATE KEY CONSTRUCTION

In this paper, binary vectors are denoted using bold lowercase letters (e.g., $\boldsymbol{a}$, $\boldsymbol{b}$) with their corresponding polynomial representations written as $\boldsymbol{a}(x)$, $\boldsymbol{b}(x)$. The notations $\boldsymbol{a}$ and $\boldsymbol{a}(x)$ are used interchangeably depending on the context. The Hamming weight of a vector $\boldsymbol{a}$ is denoted by wt($\boldsymbol{a}$). We adopt standard conventions from error-correcting code theory: $\boldsymbol{m}$ denotes an information sequence; $\boldsymbol{c}$ and $\boldsymbol{d}$ represent codewords (for masked and non-masked codes, respectively); $\boldsymbol{s}$ denotes a syndrome; and $\boldsymbol{r}$ represents a CRC polynomial. We use calligraphic letters, such as $\mathcal{D}$ and $\mathcal{L}$ to denote sets of binary vectors. Polynomial generator matrices associated with convolutional codes are denoted by uppercase letters (e.g., $A(x)$, $B(x)$), with their corresponding scalar representations written as $A$ and $B$. Given $n$ vectors:

$$\boldsymbol{v} = (v_0, v_1, v_2, \ldots), \boldsymbol{u} = (u_0, u_1, u_2, \ldots), \boldsymbol{w} = (w_0, w_1, w_2, \ldots), \ldots$$

the *interleaving* operation, denoted $\boldsymbol{u}\wedge\boldsymbol{v}\wedge\boldsymbol{w}\ldots$, produces a new vector by alternating the elements of the input vectors:

$$(\boldsymbol{v}\wedge\boldsymbol{u}\wedge\boldsymbol{w}\ldots) = (v_0, u_0, w_0, \ldots v_1, u_1, w_1, \ldots v_2, u_2, w_2, \ldots) \tag{1}$$

To *Deinterleave* an interleaved vector and extract one of its constituent components, we define the following operation:

$$(\boldsymbol{v} \curlywedge \boldsymbol{u} \curlywedge \boldsymbol{w} \ldots)_i \tag{2}$$

This operation extracts the elements located at positions

$$i,\ n+i,\ 2n+i,\ \ldots$$

in the interleaved vector. Specifically,

$$(\boldsymbol{v} \curlywedge \boldsymbol{u} \curlywedge \boldsymbol{w} \ldots)_0 = \boldsymbol{v}$$

Denote by $\boldsymbol{p}_i(x)$ a binary polynomial of memory up to $p$

$$\boldsymbol{p}_i(x) = \sum_{j=0}^{p} a_j x^j, \text{ where } a_j \in \mathbb{F}_2 \tag{3}$$

A Convolutional Code (CC) is defined by a set of $n$ constituent polynomials, customarily structured into the form of a *polynomial generator matrix,* denoted as

$$G_P(x) = [\boldsymbol{p}_0(x), \boldsymbol{p}_1(x), \ldots, \boldsymbol{p}_{n-1}(x)] \tag{4}$$

The matrix $G_P(x)$ determines both the rate and the error-correction capability of the CC, which is typically characterized by its *free distance*, $d_{free}$. The parameter $p$ determines the number of states, $2^p$, in the trellis diagram—a directed graph representing the code structure—and thus influences complexity of decoding using the Viterbi algorithm.

For simplicity, we consider good CCs of rate $1/n$, although the construction presented herein is applicable to any CC, including punctured codes. The matrix $G_P(x)$ has a scalar representation $G_P$ given by:

$$G_P = \begin{bmatrix} \boldsymbol{g}_0 & \boldsymbol{g}_1 & \boldsymbol{g}_2 & \cdots & \boldsymbol{g}_p & \boldsymbol{0} & \boldsymbol{0} & \boldsymbol{0} & \cdots \\ \boldsymbol{0} & \boldsymbol{g}_0 & \boldsymbol{g}_1 & \cdots & \boldsymbol{g}_{p-1} & \boldsymbol{g}_p & \boldsymbol{0} & \boldsymbol{0} & \cdots \\ \boldsymbol{0} & \boldsymbol{0} & \boldsymbol{g}_0 & \cdots & \boldsymbol{g}_{p-2} & \boldsymbol{g}_{p-1} & \boldsymbol{g}_p & \boldsymbol{0} & \cdots \\ \vdots & & & & & & & & \end{bmatrix} \tag{5}$$

where $\boldsymbol{g}_i$ is the $1\times n$ matrix of coefficients of $x^i$ (in the general case $\boldsymbol{g}_i$ is a $k \times n$ scalar matrix). For example, for $G_P(x) = [1+x^2,\ 1+x+x^2]$ we have $\boldsymbol{g}_0 = [1\ 1]$, $\boldsymbol{g}_1 = [0\ 1]$, $\boldsymbol{g}_2 = [1\ 1]$ and $\boldsymbol{0} = [0\ 0]$.

To each polynomial $\boldsymbol{p}_i(x)$ we associate a *high-memory polynomial* $\boldsymbol{q}_i(x)$ with degree up to $q$,

$$\boldsymbol{q}_i(x) = \sum_{j=0}^{q} b_j x^j, \text{ where } b_j \in \mathbb{F}_2 \tag{6}$$

The set of these $n$ high-memory polynomials forms the polynomial matrix:

$$G_Q(x) = [\boldsymbol{q}_0(x), \boldsymbol{q}_1(x), \ldots, \boldsymbol{q}_{n-1}(x)] \tag{7}$$

The proposed algorithm imposes no restriction on the choice of $G_P(x)$ and $G_Q(x)$, and their polynomials may be reducible or irreducible. However, in practice the degrees $p$ and $q$ are chosen such that $p + q > 200$ and $p \ll q$. Furthermore, the selection of $G_Q(x)$ significantly affects the propagation of the error during decryption (as discussed in Section 6), necessitating an appropriate choice to maintain a tolerable error rate at the decoder. The *high-memory* polynomial generator matrix $G_{PQ}(x)$ is defined as:

$$G_{PQ}(x) = [\boldsymbol{p}_0(x)\boldsymbol{q}_0(x), \boldsymbol{p}_1(x)\boldsymbol{q}_1(x), \ldots, \boldsymbol{p}_{n\text{-}1}(x)\boldsymbol{q}_{n\text{-}1}(x)] \tag{8}$$

A corresponding finite-dimensional high-memory scalar generator matrix is given by

$$G_{PQ} = \begin{bmatrix} \boldsymbol{g}_0 & \boldsymbol{g}_1 & \boldsymbol{g}_2 & \cdots & \boldsymbol{g}_{p+q} & \mathbf{0} & \cdots & & & \mathbf{0} \\ \mathbf{0} & \boldsymbol{g}_0 & \boldsymbol{g}_1 & \cdots & \boldsymbol{g}_{p+q-1} & \boldsymbol{g}_{p+q} & \mathbf{0} & \cdots & & \mathbf{0} \\ \mathbf{0} & \mathbf{0} & \boldsymbol{g}_0 & \cdots & \boldsymbol{g}_{p+q-2} & \boldsymbol{g}_{p+q-1} & \boldsymbol{g}_{p+q} & \mathbf{0} & \cdots & \mathbf{0} \\ \vdots \\ \mathbf{0} & \mathbf{0} & \mathbf{0} & \ldots & \boldsymbol{g}_0 & \boldsymbol{g}_1 & \cdots & \boldsymbol{g}_{p+q-2} & \boldsymbol{g}_{p+q-1} & \boldsymbol{g}_{p+q} \end{bmatrix} \tag{9}$$

The number $K$ of rows of $G_{PQ}$ and the corresponding number of columns, given by

$$N = n(K + p + q)$$

can be determined by the owner of the key. Multiplying each polynomial of $G_P(x)$ by a high-memory polynomial substantially increases the run-length of each row in $G_{PQ}$, i.e., the length of a sequence within the row that begins and ends with "1," compared to $G_P$.

When employed for error correction, the $K \times N$ matrix $G_{PQ}$ corresponds to a CC with memory length $p + q$, that can be described using a trellis diagram comprising $N/n$ segments and up to $2^{p+q}$ states. A conventional trellis diagram starts from a single state and expands to $2^{p+q}$ states over $p + q$ segments. Additionally, $p + q$ zeroes need be appended at end of the information sequence to drive the trellis to a single state. In principle, a Viterbi decoder can be used for maximum-likelihood hard-decision decoding (based on a Hamming distance metric) of the CC. However, this approach becomes impractical for large values of $2^{p+q}$.

Since $G_{PQ}$ has finite dimensions, the corresponding CC exhibits block code properties, defined by a generator matrix with a structured diagonal pattern. Each row is obtained by a right shift of $n$ bits relative to the previous row. Although the code description is intricate, permuting the columns of $G_{PQ}$ does not obscure its structure, as the position of zero sequences in each column still reveal discernible patterns.

To obfuscate $G_{PQ}$, disrupting its diagonal structure while maximizing the run length, we introduce a *$K \times N$ masking matrix $\tilde{G}$ of rank $l$*, where each row is randomly selected from a predefined set $\mathcal{L}$ of up to $2^l$ random binary vectors of length $N$. Notably, the set $\mathcal{L}$ can be constructed from random vectors and their linear combinations or, more conveniently, from the same $n$-tuple of Hamming weight approximately $n/2$, repeated $N/n$ times. This structured construction does not significantly reduce resistance to cryptanalysis and retains the same computational complexity as using fully random vectors. Although adopted primarily to simplify the notation, it also facilitates a clearer presentation of the decryption process. For example, when $n = 2$, there are only two such vectors:

$$\mathcal{L} = \{(101010\ldots10), (010101\ldots01)\}. \tag{10}$$

For $n = 4$, the set $\mathcal{L}$ could be taken as:

$$\begin{aligned}\mathcal{L} = \{&(11001100 \ldots 1100), (10101010 \ldots 1010), (10011001 \ldots 1001), \\ &(01100110 \ldots 0110), (01010101 \ldots 0101), (00110011 \ldots 0011)\}\end{aligned} \tag{11}$$

The masked generator matrix is then constructed as $G_{PQ} + \tilde{G}$, where addition is performed over $\mathbb{F}_2$. The resulting matrix is now fully dense, having lost its diagonal structure. Also, since $G_{PQ}$ has full rank then $G_{PQ} + \tilde{G}$ can be chosen to have full rank and is therefore well suited for subsequent invertible transformations. The final encryption matrix, which serves as the public key, is obtained by applying both row and column transformations:

$$G = S\left(G_{PQ} + \tilde{G}\right)R \tag{12}$$

where $S$ is a random non-singular $K \times K$ binary matrix used to hide the encoding—that is correspondence between plaintexts (information words) and ciphertexts (codewords). The matrix $R$ is a random $N \times N$ permutation matrix. We refer to the code described by $G$, a *high-memory masked Convolutional Code* (MCC).

The transformation from $G_P$ to $G$ involves several steps, some reversible and others semi-reversible, ultimately producing a fully dense, random-like matrix structure. This structure offers significantly stronger resistance to cryptanalysis compared to low-density matrices, or those with distinct, recognizable patterns.

Let $e$ denote the probability that the sender flips a ciphertext bit. The parameter $e$ is selected by the public key owner (i.e., the recipient) and is typically chosen to be sufficiently large to enhance resistance against cryptanalytic attacks.

To finalize the construction of the public key, an error-detection mechanism is introduced using a cyclic redundancy check (CRC), defined by a polynomial $\boldsymbol{r}(x)$ of degree $r$. This encoding step, applied to the plaintext, enables the recipient to detect decoding failures. Such detection is essential because, unlike block codes, CCs do not guarantee successful decoding, even when the error rate is within correctable bounds.

We are now ready to define the *public key* as:

$$\{G, e, \boldsymbol{r}\} \tag{13}$$

and the *private key* as:

$$\{S, R, G_P(x), G_Q(x), \tilde{G}\}. \tag{14}$$

At the decoder, the inverse permutation must be applied, and the effect of the masking matrix $\tilde{G}$ must be reversed. However, this masking operation is not entirely reversible. While the recipient has full knowledge of $\tilde{G}$, the plaintext remains unknown. Consequently, the specific linear combination of the rows of $\tilde{G}$ that was added to the ciphertext cannot be uniquely recovered. The method used to address this ambiguity is detailed in Section 4.

In addition, the obfuscation induced by the high-memory polynomials must be considered prior to decoding. A trellis decoder with feasible complexity can only operate directly on the generator matrix $G_P$ and not on its high-memory variant $G_{PQ}$. Attempting to invert multiplication by high-memory polynomials in the presence of noise is inherently difficult, as it can trigger error propagation and increase the likelihood of decryption failures at the recipient. At the same time, this very error propagation amplifies cryptanalytic complexity, thereby strengthening security. A judicious choice of $G_Q(x)$ can balance these effects, limiting error propagation while preserving the desired security margin.

## 3. ENCRYPTION BY SENDER

Assume that the sender possesses the public key. The following steps are performed by the sender to generate the ciphertext:

**Step1: Generation of Plaintext:**

A random plaintext $\boldsymbol{m}$ of length $K - r$ bits is generated, and its polynomial representation is denoted by $\boldsymbol{m}(x)$.

**Step 2: Appending CRC**

To append $r$ CRC bits to $\boldsymbol{m}$, the following procedure is used:
- Multiply $\boldsymbol{m}(x)$ by $x^r$ (effectively shifting it).
- Divide $x^r\boldsymbol{m}(x)$ by the polynomial $\boldsymbol{r}(x)$ and compute the remainder.
- Append the remainder to $\boldsymbol{m}$ to form a binary vector of length $K$, denoted as $\boldsymbol{m}_r$.

**Step 3: Codeword Generation**

The codeword $\boldsymbol{c}$ of length $N$ is calculated as:

$$\boldsymbol{c} = \boldsymbol{m}_r G. \tag{15}$$

**Step 4: Error Introduction**

Random errors are introduced to $\boldsymbol{c}$ by flipping each bit with probability $e$. The actual number of errors injected by the sender is unknown. Denote the resulting ciphertext as:

$$\boldsymbol{c}_e = \boldsymbol{c} + \boldsymbol{e} \tag{16}$$

where $\boldsymbol{e}$ is the random error vector generated by the sender.

**Step 5: Transmission**

The ciphertext $\boldsymbol{c}_e$ is then transmitted by the sender to the recipient.

Notably, since $\boldsymbol{e}$ is generated randomly, it can contain more than $eN$ errors or form localized clusters that increase the risk of decoding failure. Furthermore, due to polynomial division at the decoder, the effective error weight may increase further, contributing to additional obfuscation. Nevertheless, due to the strong error-correcting performance of the CC defined by $G_P$, the probability of decoding failure remains very low even for relatively large values of $e$, as demonstrated in Section 8. In practical implementations, any rare decoding failures are reliably detected by the CRC, allowing the recipient to either select the next most likely plaintext candidate or request a retransmission.

## 4. DECRYPTION BY THE RECIPIENT

The recipient possesses the public and private keys along with the received ciphertext, i.e.,

$$\{S, R, G_P(x), G_Q(x), \tilde{G}, G, e, \boldsymbol{r}, \boldsymbol{c}_e\} \tag{17}$$

The following steps are performed by the recipient to decrypt the ciphertext:

**Step 1: Inverse Permutation**

Apply the inverse permutation to $\boldsymbol{c}_e$ to obtain $\tilde{\boldsymbol{c}}$. (Note that if $R$ is a permutation matrix, then $R^{-1} = R^T$). Thus,

$$\tilde{\boldsymbol{c}} = \boldsymbol{c}_e R^T \quad (18)$$

Since $\boldsymbol{e}$ is a random error vector with an unknown Hamming weight,

$$\boldsymbol{e}R^T = \boldsymbol{c}_e R^T - \boldsymbol{c}R^T \quad (19)$$

is simply another random error vector with the same weight, i.e.,

$$\mathrm{wt}(\boldsymbol{e}R^T) = \mathrm{wt}(\boldsymbol{e}) \quad (20)$$

**Step 2: Unmasking**

Reverse the masking introduced by the summation of $\tilde{G}$ and $G_{PQ}$. Denote by LS($\mathcal{L}$) the linear span of the set $\mathcal{L}$. Since $\mathcal{L}$ contains *l* vectors, we have $|\mathrm{LS}(\mathcal{L})| \leq 2^l$. The vector $\tilde{\boldsymbol{c}}$ can be expressed as:

$$\tilde{\boldsymbol{c}} = \boldsymbol{m}_r S G_{PQ} + \boldsymbol{m}_r S\tilde{G} = \boldsymbol{m}_r S G_{PQ} + \boldsymbol{l}_i, \text{ with } \boldsymbol{l}_i \in \mathrm{LS}(\mathcal{L}) \quad (21)$$

However, although $\tilde{G}$ is known to the recipient, the value of $\boldsymbol{l}_i$ is unknown and may be any member of LS($\mathcal{L}$). Therefore, the set of possible unmasked vectors, denoted by $\mathcal{M}$, is given by:

$$\mathcal{M} = \{\tilde{\boldsymbol{c}} - \boldsymbol{l}_i \mid \boldsymbol{l}_i \in \mathrm{LS}(\mathcal{L})\} \quad (22)$$

For example, if the CC has rate 1/4 (i.e., *n* = 4) and the set $\mathcal{L}$ is chosen as in Equation (11), then

$$\mathrm{LS}(\mathcal{L}) = \{(0000 \ldots 0000), (1000 \ldots 1000), (0100 \ldots 0100), \ldots, (1111 \ldots 1111\} \quad (23)$$

so that $\mathcal{M}$ contains $|\mathrm{LS}(\mathcal{L})| = 16$ unmasked candidate vectors.

**Step 3: Inverting the High-Memory Polynomial Multiplication**

The next step involves reversing the multiplication by the high-memory polynomials $G_Q(x)$. Using polynomial representations, each member of $\mathcal{M}$ can be regarded as the result of interleaving the following *n* polynomials, each of length *N/n* :

$$(\tilde{\boldsymbol{c}} - \boldsymbol{l}_i)_0 \curlywedge (\tilde{\boldsymbol{c}} - \boldsymbol{l}_i)_1 \curlywedge \ldots \curlywedge (\tilde{\boldsymbol{c}} - \boldsymbol{l}_i)_{n-1} \quad (24)$$

where the elements of $(\tilde{\boldsymbol{c}} - \boldsymbol{l}_i)_j$ appear at the

$$j\text{th}, (n + j)\text{th}, (2n + j)\text{th}, \ldots, (\tfrac{N}{n} - 1 + j)\text{th}$$

positions of the interleaved vector $\tilde{\boldsymbol{c}} - \boldsymbol{l}_i$. Therefore, each such polynomial, denoted $\left(\tilde{\boldsymbol{c}}(x) - \boldsymbol{l}_i(x)\right)_j$, shall be divided by the corresponding polynomial $\boldsymbol{q}_j$ (*x*) to obtain a quotient polynomial, denoted by$\left(\boldsymbol{d}_i(x)\right)_j$. The value of $\left(\boldsymbol{d}_i(x)\right)_j$ naturally depends on the value of $\boldsymbol{l}_i(x)$.

For simplicity, assume that $\mathcal{L}$ was selected as in Equation (10). Under this condition, each polynomial $\left(\tilde{\boldsymbol{c}}(x) - \boldsymbol{l}_i(x)\right)_j$ arises from masking with either an all-zero masking polynomial $\mathbf{0}(x)$, where $\boldsymbol{l}_i(x) = \mathbf{0}(x)$, or an all-one masking polynomial $\mathbf{1}(x)$, where $\boldsymbol{l}_i(x) = \mathbf{1}(x)$. This results with two possible quotient values. when $\boldsymbol{l}_i(x) = \mathbf{0}(x)$

$$\left(\boldsymbol{d}_0(x)\right)_j = \frac{\left(\tilde{\boldsymbol{c}}(x) - \mathbf{0}(x)\right)_j}{\boldsymbol{q}_j(x)} \tag{25}$$

When $\boldsymbol{l}_i(x) = \mathbf{1}(x)$

$$\left(\boldsymbol{d}_1(x)\right)_j = \frac{\left(\tilde{\boldsymbol{c}}(x) - \mathbf{1}(x)\right)_j}{\boldsymbol{q}_j(x)} \tag{26}$$

At this stage of decryption, the recipient cannot determine which quotient, $\left(\boldsymbol{d}_0(x)\right)_j$ or $\left(\boldsymbol{d}_1(x)\right)_j$, is the correct one for index $j$. Furthermore, since $\left(\tilde{\boldsymbol{c}}(x) - \boldsymbol{l}_i(x)\right)_j$ may contain errors, the division might yield non-zero remainders. These remainders can either be ignored or be detected and subtracted if $\boldsymbol{q}_j(x)$ is carefully chosen to function as an error-detecting code. When ignored, we assume that errors left in the quotient will be corrected by the Viterbi decoder in Step 5 below.

**Step 4: Interleaving of Quotients**

The quotients are re-interleaved to form a vector $\boldsymbol{d}_i$:

$$\boldsymbol{d}_i = (\boldsymbol{d}_i)_0 \curlywedge (\boldsymbol{d}_i)_1 \curlywedge \dots \tag{27}$$

In the general case, there are no more than $2^l$ variants in the set of interleaved quotients, denoted as

$$\mathfrak{D} = \{\boldsymbol{d}_i\}_{i=0}^{2^l-1} \tag{28}$$

All candidates in $\mathfrak{D}$ need be considered.

**Step 5: Parallel Viterbi Decoding**

We now aim to determine $\widehat{\boldsymbol{m}}_r$, the most likely value of $\boldsymbol{m}_r$ (i.e., the plaintext with the appended CRC), by applying Viterbi decoding to all possible members of the set $\mathfrak{D}$. Since $\mathfrak{D}$ contains $2^l$ candidate vectors, these decodings can be performed in parallel for improved efficiency.

The most likely codeword $\widehat{\boldsymbol{d}}$ is determined by

$$\widehat{\boldsymbol{d}} = \boldsymbol{d}_i - \widehat{\boldsymbol{e}} \tag{29}$$

where $\widehat{\boldsymbol{e}}$ is the error vector with the minimum Hamming weight among the outcomes of all $2^l$ parallel Viterbi decoders. The index $i$ in Equation (29) corresponds to the decoder that processes the correct $\boldsymbol{d}_i$. The same Viterbi decoder also reveals $\widehat{\boldsymbol{m}}_r S$, the transformed plaintext with appended CRC that generated $\widehat{\boldsymbol{d}}$.

Importantly, a key distinction exists among the $2^l$ members of $\mathfrak{D}$. For exactly one variant, say $\boldsymbol{d}_i$, Viterbi decoding will yield a codeword at a Hamming distance of approximately $eN + \alpha$ from $\boldsymbol{d}_i$, where $eN$ is the injected error weight, and $\alpha$ denotes the number of additional errors introduced by the polynomial division process. The value of $\alpha$ can be estimated via simulations, as described in Section 6. By contrast, decoding any of the other $2^l$ -1 candidates in $\mathfrak{D}$ typically produces codewords at substantially larger Hamming distance than $eN + \alpha$. This larger Hamming distance can be estimated from the CC's weight spectrum, using the Gilbert bound analysis presented in Section 6.

This distinction arises because subtracting an incorrect masking vector from $\tilde{\boldsymbol{c}}$ effectively introduces random errors into the candidate $\boldsymbol{d}_i$, allowing the decoder to reliably identify the correct candidate among all alternatives

**Step 6: Plaintext Recovery**

To recover the original plaintext, we reverse the transformation induced by the matrix $S$:

$$\widehat{\boldsymbol{m}}_r = \widehat{\boldsymbol{m}}_r S S^{-1} \tag{30}$$

Next, the remainder obtained from dividing $\widehat{\boldsymbol{m}}_r(x)$ by $\boldsymbol{r}(x)$ must be computed. If this remainder is zero, we declare that

$$\boldsymbol{m}_r = \widehat{\boldsymbol{m}}_r \tag{31}$$

and recover the plaintext $\boldsymbol{m}$ by discarding the $r$ CRC bits from $\boldsymbol{m}_r$. If the remainder is nonzero, the selected codeword $\widehat{\boldsymbol{d}}$ is rejected, and the process continues iteratively with the next most likely candidate. This process is repeated until a valid plaintext is identified or until all candidates are exhausted. If no valid plaintext is found, a retransmission of another ciphertext is requested.

Finally, we remark that the structure of the matrix $G_P$ corresponds to a trellis that terminates in a single state. As a result, the Viterbi decoder returns the most likely information sequence padded with $p$ trailing zeros. To recover the original plaintext, theses appended zeroes must be removed from the decoded output. This step ensures that the final recovered message accurately reflects the original input.

The encryption and decryption algorithms are illustrated in the block diagram of Fig. 1.

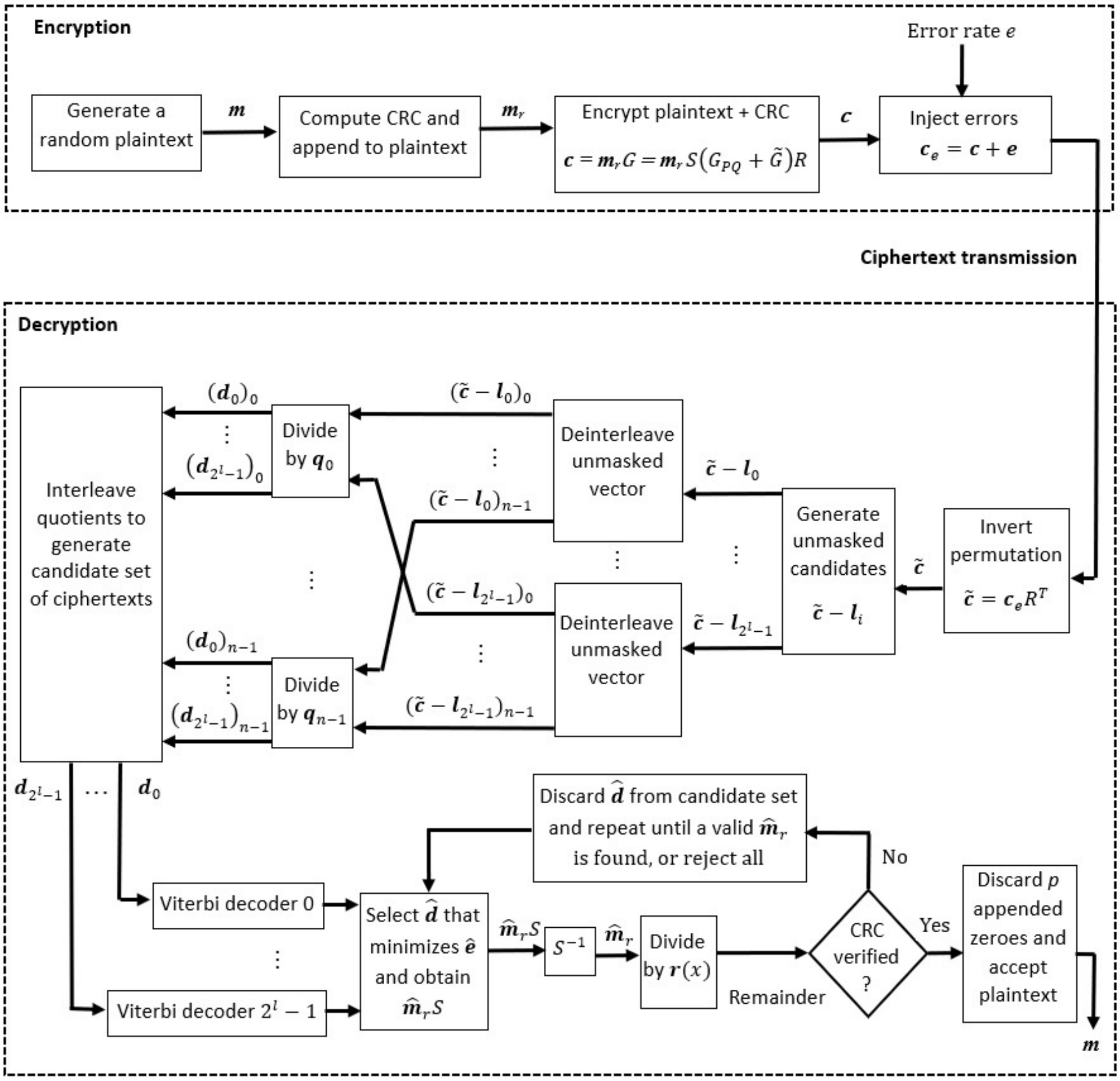

Figure 1. Encryption and Decryption Block Diagram

## 5. SECURITY AGAINST STRUCTURAL ATTACKS

An eavesdropper observes the public key $G$ and the intercepted ciphertext $\boldsymbol{c_e}$. However, since $\boldsymbol{c_e}$ includes injected errors, inverting $G$ directly is useless for recovering the plaintext $\boldsymbol{m}$. The attacker's options are therefore limited to either (i) enumerating all possible error patterns—an operation exponential in the error weight—or (ii) attempting to recover the structured matrix $G_{PQ}$ from the public key $G$. The second approach is defeated by three independent obfuscation layers:

- **Random column permutation** $R$, which scrambles all coordinate positions.
- **Random nonsingular** $S$, which changes the basis of the code.
- **Dense masking** $\tilde{G}$, a low-rank matrix with random rows of weight close to *N/2*, carefully constructed so that each row and column is statistically balanced.

**Rank preservation:** Observing the structure of $G = S(G_{PQ} + \tilde{G})R$, with $G_{PQ} \in \mathbb{F}_2^{K\times N}$ of full row rank *K*, and $\tilde{G}$ a rank-*l* dense masking matrix. Hence, the sum $G_{PQ} + \tilde{G}$ has rank at least *K* - *l*.

Because the fixed $l$ row patterns of $\tilde{G}$ are dense and mutually distinct, the sum will in general have full rank $K$. Moreover, knowing the diagonal structure of $G_{PQ}$ allows the selection of random row vectors in $\tilde{G}$ to guarantee this property. Multiplication by $S$ and permutation by $R$ preserve rank, so the public key $G$ remains a full-rank generator matrix.

**Column randomness via Linear Tests:** Denote by $y_j$ the $j$th column of $S\left(G_{PQ} + \tilde{G}\right)$. Because the rows of $G_{PQ} + \tilde{G}$ are dense with Hamming weight $\approx N/2$ and are subsequently scrambled by a random nonsingular matrix $S$, the entries of $y_j$ behave like independent Bernoulli(1/2) bits. Consequently, the column weights of $S\left(G_{PQ} + \tilde{G}\right)$ concentrate sharply around $K/2$, as expected in a uniformly random full-rank code.

A *linear test* is defined by a nonzero vector $a \in \mathbb{F}_2^K$, producing the parity $a^{\mathrm{T}} y_j \in \{0,1\}$. This parity is 0 if the overlap between $a$ and $y_j$ has even Hamming weight and 1 otherwise. Any persistent bias would require alignment between $a$ and specific row patterns. However, multiplication by $S$ spreads each fixed row-pattern across all coordinates, eliminating such alignments.

Thus, the outcomes of linear tests are essentially uniform, and the empirical column statistics are indistinguishable from those of a full-rank random $(N, K)$ block code, up to deviations on the order of $2^{-l}$ (set by the number of fixed patterns). For practical values of $l$, this deviation is negligible compared with natural sampling fluctuations.

**Suppression of Low-Weight Duals:** A vector $w \in \mathbb{F}_2^K \setminus \{0\}$ is dual if

$$w^T\left(G_{PQ} + \tilde{G}\right) = 0.$$

Since $G_{PQ}$ has full rank, $w^T G_{PQ}$ is nonzero for all $w \neq 0$. The equation $w^T G_{PQ} = w^T \tilde{G}$ requires that combination of fixed row patterns specified by $w$ exactly reproduce the nonzero target vector $w^T G_{PQ}$. For dense, mutually distinct patterns, the probability of such alignment for any low-weight $w$ is negligible. After multiplication by $S$ and permutation $R$, any such rare alignment is uncorrelated with specific coordinate positions. Consequently, the number and distribution of low-weight duals matches that of a random $(N, K)$ code.

**Masking Entropy:** A key asymmetry arises from the masking matrix $\tilde{G}$: for the adversary, $\tilde{G}$ is completely unknown. Although its rank $l$ is small, the space of possible masks is enormous. Each of the $K$ rows can be chosen independently from $l$ fixed patterns (or their linear combinations), yielding at least $l^K$ distinct possible masks. For instance, with $K = 2000$ and $l = 5$, there are $5^{2000}$ candidates, corresponding to about 4644 bits of entropy. Exhaustive search over this space is infeasible in practice, and ciphertext errors prevent efficient verification of guessed masks.

By contrast, the legitimate recipient who constructed $\tilde{G}$ only needs to check $2^l$ mask candidates, and can perform polynomial-time parallel decoding, selecting the codeword with smallest Hamming distance. This exponential-versus-polynomial gap is a fundamental source of the security argument.

In conclusion, the obfuscations $S\left(G_{PQ} + \tilde{G}\right)R$ erase visible patterns in $G_P$ and make the columns of $G$ statistically indistinguishable from those of a uniformly random $(N, K)$ block code. Weight distributions, low-weight dual codewords, and other algebraic fingerprints that could reveal the underlying structure are no longer detectable. Given the injected ciphertext errors, the only known generic decoding method is Information-Set Decoding (ISD), with complexity exponential in the error weight.

**Attack from a Known Convolutional Code:** Even if the attacker knew the exact underlying high-memory CC used to generate $G_{PQ}$, direct decoding would remain infeasible. In the public construction, the code's generator matrix $G_P$ is multiplied by high-degree polynomials (smearing the run-length), and then corrupted with errors. The private decoder can invert these polynomials, reducing the problem to a compact trellis, but in doing so also spreads the errors in a controlled way that only the intended decoding process can handle. Without undoing the permutation $R$, removing $\tilde{G}$, and reversing the polynomial multiplication— while contending with error propagation—no trellis-based attack is possible. In practice, these layers leave the adversary facing what is effectively a random code, for which ISD remains the sole feasible decoding strategy.

## 6. Polynomial Selection and Expected Error Rate

The decryption algorithm involves polynomial divisions that can cause error propagation; a single error in the polynomials $\left(\tilde{\boldsymbol{c}}(x) - \boldsymbol{l}_i(x)\right)_j$ may lead to multiple errors in the quotient. To reduce this risk, the polynomial matrix $G_P(x)$ is selected to ensure that the corresponding CC has strong error-correction capabilities (e.g., with free distance $d_{\text{free}} > 20$), outperforming Goppa codes or similar linear block codes with comparable size. This results in rare decoding failures, even at relatively high error rates $e$ (a larger $e$ also implies improved security).

To further limit the probability of error propagation, the polynomials $G_Q(x)$ should be chosen to limit the spread of isolated errors in $\left(\tilde{\boldsymbol{c}}(x) - \boldsymbol{l}_i(x)\right)_j$, confining them to just a few errors in the quotient. Given an error rate $e$, simulation can test $G_Q(x)$ by measuring the number of quotient errors for various random error vectors with the same $e$, allowing for an informed selection of the CC.

Let $\boldsymbol{e}_j$ be a random error vector of length $N/n$ with error rate $e$. The total number of additional errors introduced by polynomial division, denoted α, is given by:

$$\alpha = \sum_{j=0}^{n-1} \left[\mathrm{wt}\left(\frac{\boldsymbol{e}_j}{\boldsymbol{q}_j}\right) - \mathrm{wt}(\boldsymbol{e}_j)\right] \tag{32}$$

Here, the term $\frac{\boldsymbol{e}_j}{\boldsymbol{q}_j}$ corresponds to the quotient obtained during polynomial division of an error vector of length $N/n$ by the divisor polynomial. The expression in Equation (32) captures the cumulative increase in Hamming weight due to the spread of errors in the quotient domain.

A practical way to limit the spread of isolated errors during polynomial division is to choose $\boldsymbol{q}_j$ as a sparse polynomial with widely spaced nonzero exponents. For instance, the trivial choice $\boldsymbol{q}_j(x) = x^i$ (with ($i \leq q$), completely avoids error propagation, although it contributes minimally to security. A more effective option is to use a two-term polynomial, which both extends the run-length of each row while minimizing error spread:

$$\boldsymbol{q}_j(x) = 1 + x^q \tag{33}$$

Increasing the number of nonzero elements in $\boldsymbol{q}_j(x)$ can dramatically amplify the difference:

$$\mathrm{wt}\left(\frac{\boldsymbol{e}_j}{\boldsymbol{q}_j}\right) - \mathrm{wt}(\boldsymbol{e}_j) \tag{34}$$

Therefore, the polynomials of $G_Q(x)$ should be selected such that the resulting estimated error rate

$$\frac{eN+\alpha}{N} \tag{35}$$

remains within the CC's decoding capacity.

Consider a candidate vector $\boldsymbol{d}_i$ defined according to Equation (27), and assume that $\boldsymbol{d}_i$ is an incorrect candidate (i.e., it was incorrectly demasked). We now estimate the Hamming distance between $\boldsymbol{d}_i$ and the closest codeword in the CC. Maximum-likelihood hard-decision Viterbi decoding of a random vector such as $\boldsymbol{d}_i$ effectively returns the coset leader—the error vector that produces the nearest codeword in Hamming distance.

The expected weight of a random coset leader can be approximated using the Gilbert bound, which provides a good estimate of the number of errors corrected in a random vector. For essentially any "good" binary linear code of rate $\rho = K/N$, the typical nearest-neighbour distance $\delta$ is well approximated by the solution of the volume (Gilbert) equation

$$\sum_{i=0}^{\delta} \binom{N}{i} \approx 2^{N(1-\rho)}$$

Using the entropy approximation

$$\sum_{i\le\delta} \binom{N}{i} \approx 2^{NE(\delta/N)}$$

we obtain

$$E(\delta/N) \approx 1 - \rho$$

where is $E(\cdot)$ is the binary entropy function (base-2). For example, at $\rho = \frac{1}{2}$ we have

$$E\left(\frac{\delta}{N}\right) \approx \frac{1}{2} \Rightarrow \frac{\delta}{N} \approx 0.11$$

Similarly, at code rates $\rho = \frac{1}{3}$ and $\rho = \frac{1}{4}$ we obtain $\frac{\delta}{N} \approx 0.174$ and $\frac{\delta}{N} \approx 0.215$, respectively.

## 7. WORKED EXAMPLE

Define the following CC, $G_P(x)$ = $[\boldsymbol{p}_0(x),\ \boldsymbol{p}_1(x)]$ = $[1+x^2,\ 1+x+x^2]$. Suppose that a randomly selected plaintext of length 6 is given by $\boldsymbol{m}$ = [1 1 1 0 0 1], in polynomial form, $\boldsymbol{m}(x)$ = $1+x+x^2+x^5$. In this example we skip the trivial step of CRC construction, assuming it is already contained within $\boldsymbol{m}(x)$.

A codeword $\boldsymbol{d}(x)$ of the CC is obtained by either interleaving $\boldsymbol{m}(x)\boldsymbol{p}_0(x)$ with $\boldsymbol{m}(x)\boldsymbol{p}_1(x)$ or simply by employing a scalar generator matrix with 6 rows (corresponding to the length of $\boldsymbol{m}$) constructed according to Equation (5)

$$G_P = \begin{bmatrix} 1&1&0&1&1&1&0&0&0&0&0&0&0&0&0&0 \\ 0&0&1&1&0&1&1&1&0&0&0&0&0&0&0&0 \\ 0&0&0&0&1&1&0&1&1&1&0&0&0&0&0&0 \\ 0&0&0&0&0&0&1&1&0&1&1&1&0&0&0&0 \\ 0&0&0&0&0&0&0&0&1&1&0&1&1&1&0&0 \\ 0&0&0&0&0&0&0&0&0&0&1&1&0&1&1&1 \end{bmatrix}$$

The codeword $\boldsymbol{d}$ in vector form is given by

$$\boldsymbol{d} = \boldsymbol{m}G_P = [1\ 1\ 1\ 0\ 0\ 1]G_P = [1\ 1\ 1\ 0\ 0\ 1\ 1\ 0\ 1\ 1\ 1\ 1\ 0\ 1\ 1\ 1]$$

(which matches the interleaved polynomial-wise result). Next, choosing the polynomial matrix $G_Q(x) = [1+x^7, x^7]$, we obtain the corresponding high-memory generator matrix

$$G_{PQ}(x) = [\boldsymbol{p}_0(x)\boldsymbol{q}_0(x), \boldsymbol{p}_1(x)\boldsymbol{q}_1(x)] = [(1+x^2)(1+x^7), (1+x+x^2)(x^7)]$$

$$= [1+x^2+x^7+x^9, x^7+x^8+x^9].$$

Using $G_{PQ}(x)$, we construct the high-memory scalar generator matrix $G_{PQ}$ as in Equation (9) with:

$$\boldsymbol{g}_0 = [10], \boldsymbol{g}_1 = [00], \boldsymbol{g}_2 = [10], \boldsymbol{g}_3 = [00], \boldsymbol{g}_4 = [00], \boldsymbol{g}_5 = [00],$$

$$\boldsymbol{g}_6 = [00], \boldsymbol{g}_7 = [11], \boldsymbol{g}_8 = [01], \boldsymbol{g}_9 = [11], \ \boldsymbol{0} = [00].$$

The resulting matrix is:

$$G_{PQ} = \begin{bmatrix} 1&0&0&0&1&0&0&0&0&0&0&0&0&0&1&1&0&1&1&1&0&0&0&0&0&0&0&0&0&0 \\ 0&0&1&0&0&0&1&0&0&0&0&0&0&0&0&0&1&1&0&1&1&1&0&0&0&0&0&0&0&0 \\ 0&0&0&0&1&0&0&0&1&0&0&0&0&0&0&0&0&0&1&1&0&1&1&1&0&0&0&0&0&0 \\ 0&0&0&0&0&0&1&0&0&0&1&0&0&0&0&0&0&0&0&0&1&1&0&1&1&1&0&0&0&0 \\ 0&0&0&0&0&0&0&0&1&0&0&0&1&0&0&0&0&0&0&0&0&0&1&1&0&1&1&1&0&0 \\ 0&0&0&0&0&0&0&0&0&0&1&0&0&0&1&0&0&0&0&0&0&0&0&0&1&1&0&1&1&1 \end{bmatrix}.$$

Note that the run-length of each row has increased from 6 to 20. Construct $\tilde{G}$ by randomly choosing its rows from the set $\mathcal{L} = \{(101010\ldots10), (010101\ldots01)\}$ (or any other set of $l$ random vectors of length 30):

$$\tilde{G} = \begin{bmatrix} 1&0&1&0&1&0&1&0&1&0&1&0&1&0&1&0&1&0&1&0&1&0&1&0&1&0&1&0&1&0 \\ 0&1&0&1&0&1&0&1&0&1&0&1&0&1&0&1&0&1&0&1&0&1&0&1&0&1&0&1&0&1 \\ 1&0&1&0&1&0&1&0&1&0&1&0&1&0&1&0&1&0&1&0&1&0&1&0&1&0&1&0&1&0 \\ 0&1&0&1&0&1&0&1&0&1&0&1&0&1&0&1&0&1&0&1&0&1&0&1&0&1&0&1&0&1 \\ 0&1&0&1&0&1&0&1&0&1&0&1&0&1&0&1&0&1&0&1&0&1&0&1&0&1&0&1&0&1 \\ 1&0&1&0&1&0&1&0&1&0&1&0&1&0&1&0&1&0&1&0&1&0&1&0&1&0&1&0&1&0 \end{bmatrix}.$$

Now sum $\tilde{G}$ and $G_{PQ}$ to obtain

$$G_{PQ} + \tilde{G} = \begin{bmatrix} 0&0&1&0&0&0&1&0&1&0&1&0&1&0&0&1&1&1&0&1&1&0&1&0&1&0&1&0&1&0 \\ 0&1&1&1&0&1&1&1&0&1&0&1&0&1&0&1&1&0&0&0&1&0&0&1&0&1&0&1&0&1 \\ 1&0&1&0&0&0&1&0&0&0&1&0&1&0&1&0&1&0&0&1&1&1&0&1&1&0&1&0&1&0 \\ 0&1&0&1&0&1&1&1&0&1&1&1&0&1&0&1&0&1&0&1&1&0&0&0&1&0&0&1&0&1 \\ 0&1&0&1&0&1&0&1&1&1&0&1&1&1&0&1&0&1&0&1&0&1&1&0&0&0&1&0&0&1 \\ 1&0&1&0&1&0&1&0&1&0&0&0&1&0&0&0&1&0&1&0&1&0&1&0&0&1&1&1&0&1 \end{bmatrix}$$

which is now a high-density matrix with nearly half of the entries being "1". Multiplying $G_{PQ} + \tilde{G}$ on the left by the following nonsingular matrix $S$:

$$S = \begin{bmatrix} 1&0&0&1&0&0 \\ 0&1&0&0&0&1 \\ 0&0&1&0&0&0 \\ 0&0&1&1&1&0 \\ 0&0&0&0&1&0 \\ 0&0&1&0&1&1 \end{bmatrix},$$

we have

$$S\big(G_{PQ} + \tilde{G}\big) = \begin{bmatrix} 0&1&1&1&0&1&0&1&1&1&0&1&1&1&0&0&1&0&0&0&0&0&1&0&0&0&1&1&1&1 \\ 1&1&0&1&1&1&0&1&1&1&0&1&1&1&0&1&0&0&1&0&0&0&1&1&0&0&1&0&0&0 \\ 1&0&1&0&0&0&1&0&0&0&1&0&1&0&1&0&1&0&0&1&1&1&0&1&1&0&1&0&1&0 \\ 1&0&1&0&0&0&0&0&1&0&0&0&0&0&1&0&1&0&0&1&0&0&1&1&0&0&0&1&1&0 \\ 0&1&0&1&0&1&0&1&1&1&0&1&1&1&0&1&0&1&0&1&0&1&1&0&0&0&1&0&0&1 \\ 0&1&0&1&1&1&0&1&0&1&1&1&1&1&1&1&0&1&1&0&0&0&0&1&1&1&1&1&1&0 \end{bmatrix}.$$

Apply to $S\big(G_{PQ} + \tilde{G}\big)$ the following column permutation $\pi$, defined by the bijection function

$\pi$: (14 25 9 18 30 8 21 1 10 29 5 26 3 11 23 28 15 2 7 12 20 6 17 4 27 16 24 13 22 19)
→ (1 2 3 4 5 6 7 8 9 10 11 12 13 14 15 16 17 18 19 20 21 22 23 24 25 26 27 28 29 30)
This permutation can also be performed by multiplying $S\big(G_{PQ} + \tilde{G}\big)$ on the right by an equivalent 30×30 permutation matrix $R$:

$$G = S\big(G_{PQ} + \tilde{G}\big)R = \begin{bmatrix} 1&0&1&0&1&1&0&0&1&1&0&0&1&0&1&1&0&1&0&1&0&1&1&1&1&0&0&1&0&0 \\ 1&0&1&0&0&1&0&1&1&0&1&0&0&0&1&0&0&1&0&1&0&1&0&1&1&1&1&1&0&1 \\ 0&1&0&0&0&0&1&1&0&1&0&0&1&1&0&0&1&0&1&0&1&0&1&0&1&0&1&1&1&0 \\ 0&0&1&0&0&0&0&1&0&1&0&0&1&0&1&1&1&0&0&0&1&0&1&0&0&0&1&0&0&0 \\ 1&0&1&1&1&1&0&0&1&0&0&0&0&0&1&0&0&1&0&1&1&1&0&1&1&1&0&1&1&0 \\ 1&1&0&1&0&1&0&0&1&1&1&1&0&1&0&1&1&1&0&1&0&1&0&1&1&1&1&1&0&1 \end{bmatrix}.$$

The codeword $\boldsymbol{c}$ of the MCC corresponding to $G$ is given by

$$\boldsymbol{c} = \boldsymbol{m}G = [1\ 1\ 1\ 0\ 0\ 1]\ G = [1\ 0\ 0\ 1\ 1\ 1\ 1\ 0\ 1\ 1\ 0\ 1\ 0\ 0\ 0\ 0\ 0\ 1\ 1\ 1\ 1\ 1\ 0\ 1\ 0\ 0\ 1\ 0\ 1\ 0].$$

If we assume a randomly generated error vector $\boldsymbol{e}$ with three errors:

$$\boldsymbol{e} = [0\ 0\ 0\ 1\ 0\ 0\ 0\ 0\ 0\ 0\ 0\ 0\ 0\ 0\ 0\ 0\ 0\ 1\ 0\ 1\ 0\ 0\ 0\ 0\ 0\ 0\ 0\ 0\ 0\ 0],$$

then the received vector is:

$$\boldsymbol{c}_e = \boldsymbol{c} + \boldsymbol{e} = [1\ 0\ 0\ \underline{0}\ 1\ 1\ 1\ 0\ 1\ 1\ 0\ 1\ 0\ 0\ 0\ 0\ 0\ \underline{1}\ 1\ \underline{0}\ 1\ 1\ 1\ 0\ 1\ 0\ 0\ 1\ 0\ 1\ 0],$$

where the erroneous bits are underlined.

**Decryption**

## Step1: Inverse Permutation

Apply the inverse permutation $\pi^{-1}$ on the bits of $\boldsymbol{c}_e$:

$\pi^{-1}$:(8 18 13 24 11 22 19 6 3 9 14 20 28 1 17 26 23 4 30 21 7 29 15 27 2 12 25 16 10 5)
→ (1 2 3 4 5 6 7 8 9 10 11 12 13 14 15 16 17 18 19 20 21 22 23 24 25 26 27 28 29 30)

This operation can also be described by multiplying $\boldsymbol{c}_e$ on the right by an equivalent 30×30 permutation matrix $R^{-1}$:

$$\tilde{\boldsymbol{c}} = \boldsymbol{c}_e R^{-1} = [0\ 1\ 0\ 1\ 0\ 1\ \underline{0}\ 1\ 0\ 1\ 0\ 1\ 0\ 1\ \underline{1}\ 0\ 0\ \underline{0}\ 0\ 1\ 1\ 1\ 0\ 1\ 0\ 1\ 0\ 0\ 1\ 1].$$

## Step 2: Unmasking

The set $\mathcal{M}$ of unmasked vectors is given by Equation (22) where

$$\text{LS}(\mathcal{L}) = \{(0000 \ldots 00), (0101\ldots01), (1010 \ldots 10), \ldots, (1111 \ldots 11\}$$

We now deinterleave the four vectors of $\mathcal{M}$ to their unmasked polynomial constitutes. There are only four possible unmasked variants:

$\left(\tilde{\boldsymbol{c}}(x) - \boldsymbol{0}(x)\right)_0 = x^7+x^{10}+x^{14}$
$\left(\tilde{\boldsymbol{c}}(x) - \boldsymbol{1}(x)\right)_0 = 1+x+x^2+x^3+x^4+x^5+x^6+x^8+x^9+x^{11}+x^{12}+x^{13}$
$\left(\tilde{\boldsymbol{c}}(x) - \boldsymbol{0}(x)\right)_1 = 1+x+x^2+x^3+x^4+x^5+x^6+x^9+x^{10}+x^{11}+x^{12}+x^{14}$
$\left(\tilde{\boldsymbol{c}}(x) - \boldsymbol{1}(x)\right)_1 = x^7+x^8+x^{13}$

### Step 3: Inverting the High-Memory Polynomial Multiplication

Using Equation (25) and (26) we divide the first two outcomes of Step 2 by $\boldsymbol{q}_0(x)$ and the last two by $\boldsymbol{q}_1(x)$:

$\left(\boldsymbol{d}_0(x)\right)_0 = (x^7+x^{10}+x^{14})/(1+x^7) = x^3+x^7$ [remainder: $x^3$].
$\left(\boldsymbol{d}_1(x)\right)_0 = (1+x+x^2+x^3+x^4+x^5+x^6+x^8+x^9+x^{11}+x^{12}+x^{13})/(1+x^7)$
$= x+x^2+x^4+x^5+x^6$ [remainder: $1+x^3$].
$\left(\boldsymbol{d}_0(x)\right)_1 = (1+x+x^2+x^3+x^4+x^5+x^6+x^9+x^{10}+x^{11}+x^{12}+x^{14})/x^7$
$= x^2+x^3+x^4+x^5+x^7$ [remainder: $1+x+x^2+x^3+x^4+x^5+x^6$].
$\left(\boldsymbol{d}_1(x)\right)_1 = (x^7+x^8+x^{13})/x^7 = 1+x+x^6$ [remainder: 0].

In vector form:

$(\boldsymbol{d}_0)_0 =$ [0 0 0 1 0 0 0 1]
$(\boldsymbol{d}_1)_0 =$ [0 1 1 0 1 1 1 0]
$(\boldsymbol{d}_0)_1 =$ [0 0 1 1 1 1 0 1]
$(\boldsymbol{d}_1)_1 =$ [1 1 0 0 0 0 1 0].

## Step 4: Quotient Interleaving

The set $\mathfrak{D}$ four interleaved quotients is computed according to Equation (27):

$\boldsymbol{d}_0 = (\boldsymbol{d}_0)_0 \wedge (\boldsymbol{d}_0)_1 =$ [0 0 0 0 0 1 1 1 0 1 0 1 0 0 1 1]
$\boldsymbol{d}_1 = (\boldsymbol{d}_0)_0 \wedge (\boldsymbol{d}_1)_1 =$ [0 1 0 1 0 0 1 0 0 0 0 0 0 1 1 0]
$\boldsymbol{d}_2 = (\boldsymbol{d}_1)_0 \wedge (\boldsymbol{d}_0)_1 =$ [0 0 1 0 1 1 0 1 1 1 1 1 1 0 0 1]
$\boldsymbol{d}_3 = (\boldsymbol{d}_1)_0 \wedge (\boldsymbol{d}_1)_1 =$ [0 1 1 1 1 0 0 0 1 0 1 0 1 1 0 0]

**Step 5:** Parallel **Viterbi Decoding**

Decoding the four interleaved quotients in parallel yields the most likely codeword $\widehat{\boldsymbol{d}}$ according to Equation (30), with the corresponding transformed plaintext $\widehat{\boldsymbol{m}}_r S$. In this example, the CC has a free distance of $d_{free} = 5$, which implies that the code can reliably correct up to two errors within a sliding window of six bits. If three or more errors occur within this window, the parallel Viterbi decoding step is likely to fail in recovering the correct plaintext. Nevertheless, such decoding failure, can be detected by the CRC.

The trellis for the case $\boldsymbol{d}_3 = (\boldsymbol{d}_1)_0 \wedge (\boldsymbol{d}_1)_1$ is depicted in Fig. 2. The highlighted path in the trellis corresponding to the least-weight path has an accumulated distance of 2 from $\boldsymbol{d}_3$, which corresponds to the following information word: 11011000. For all other Viterbi decoders, the most likely path through their respective trellises maintain a minimum distance greater than two from the corresponding $\boldsymbol{d}_i$.

Note that for a CC with memory length 2, the final two zeroes result from forcing the trellis to converge to a single termination state. By discarding these appended zeroes, the most likely plaintext is obtained as:

$$\widehat{\boldsymbol{m}}S = 110110.$$

**Step 6:** Plaintext **Recovery**

Using the inverse matrix $S^{-1}$:

$$S^{-1} = \begin{bmatrix} 1&0&1&1&1&0 \\ 0&1&1&0&1&1 \\ 0&0&1&0&0&0 \\ 0&0&1&1&1&0 \\ 0&0&0&0&1&0 \\ 0&0&1&0&1&1 \end{bmatrix},$$

we can recover the original plaintext $\boldsymbol{m}$ as follows:

$$\boldsymbol{m} = \widehat{\boldsymbol{m}}SS^{-1} = [110110]S^{-1} = [111001]$$

Thus, the recovered plaintext is 111001, confirming correctness. In this example, the recovery of $\boldsymbol{m}$ required only a single iteration. In the general case, the remainder resulting from the division of $\widehat{\boldsymbol{m}}_r(x)$ by $\boldsymbol{r}(x)$ must be computed. If this remainder in non-zero, the selected codeword $\widehat{\boldsymbol{d}}$ should be discarded. The process must then be repeated iteratively with the next most likely candidate until a valid $\boldsymbol{d}_i$ is identified or all possible candidates have been exhausted.

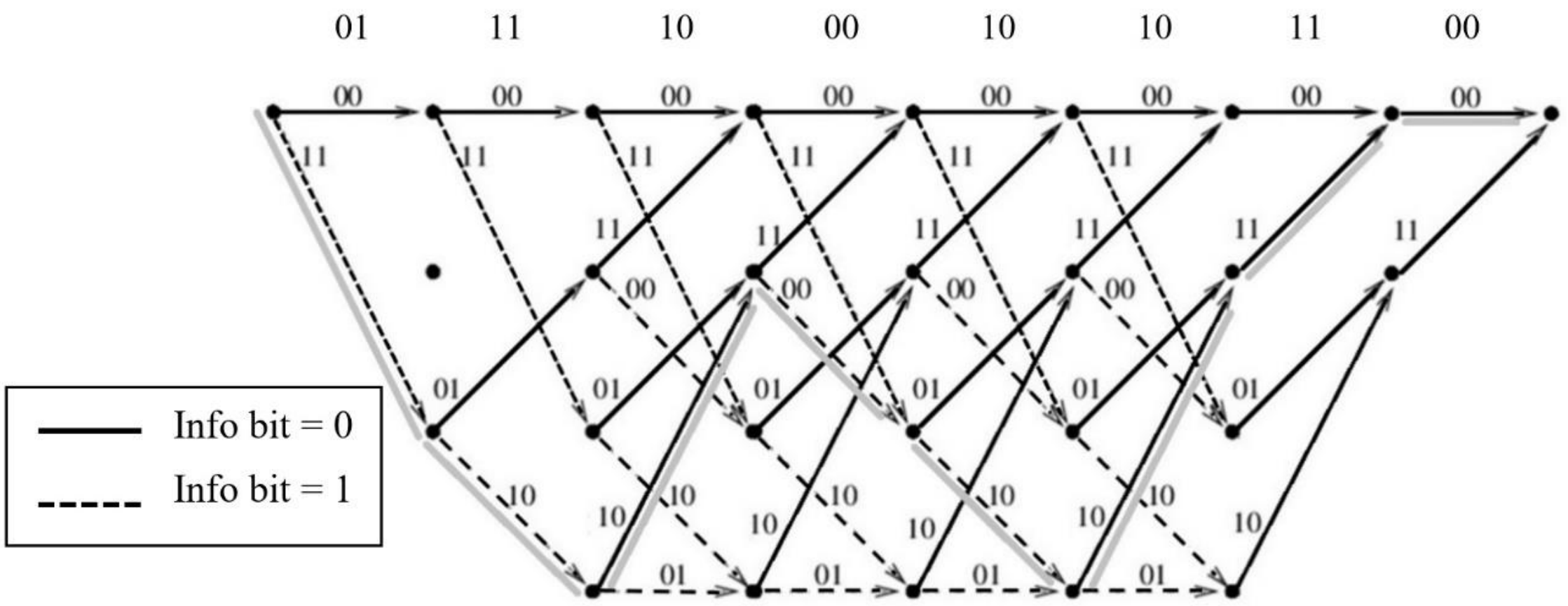

Figure 2. Trellis and least-weight path corresponding to $\boldsymbol{d}_3 = [0\ 1\ 1\ 1\ 1\ 0\ 0\ 0\ 1\ 0\ 1\ 0\ 1\ 1\ 0\ 0]$.

## 8. CRYPTANALYSIS RESISTANCE AND PERFORMANCE

In code-based cryptography, the complexity of cryptanalysis is primarily determined by the ISD algorithm, which targets the problem of decoding random linear block codes in the presence of errors. This task is considered computationally hard for large code dimensions and high error weights. However, if the public code deviates from a random code due to hidden structures, ISD may not reflect the true attack complexity. The underlying mathematical problem is the syndrome decoding problem. In our context, the finite-dimensional matrix $G$, which serves as the generator matrix for the MCC (allowing it to be treated as a linear block code), has an associated parity-check matrix $H$ satisfying:

$$GH^T = \mathbf{0} \tag{36}$$

Given a syndrome $\boldsymbol{s}$, and assuming that exactly $t$ errors have occurred, the challenge is to find an error vector $\boldsymbol{e}$ such that

$$H\boldsymbol{e}^T = \boldsymbol{s} \text{ and } \mathrm{wt}(\boldsymbol{e}) = t \tag{37}$$

The cryptanalysis complexity is estimated using the best-known ISD algorithms (e.g., Prange, Lee-Brickell, BJMM) [8-10], which assess security as bit operations, often ignoring memory usage or parallelization. These estimates assume a fixed weight $\mathrm{wt}(\boldsymbol{e})$. However, in our method, the exact error count is unknown—only the error probability $e$ is known—introducing additional complexity compared to other code-based schemes. Furthermore, while structured error patterns may affect complexity, the MCC's error vector $\boldsymbol{e}$ is truly random and, by definition, unstructured.

In the context of quantum computing, ISD complexity is typically estimated based on classical assumptions. Although quantum algorithms like Grover's [11] can offer quadratic speedup, no exponential improvement in ISD is known. Therefore, ISD-based complexity assessments in code-based cryptography assume ISD remains the most efficient attack.

To compare the complexity of MCC with other code-based methods, we adopt the complexity measure described in [11]. Specifically, for a classical computer attacking of an ($N$, $K$, $t$) code, the ISD complexity—denoted $C_{\text{ISD}}$—is proportional to the number of iterations performed by the ISD algorithm. In our notation, we have

$$C_{\text{ISD}} \sim \frac{\binom{N}{K}}{0.29\binom{N-t}{K}} \tag{38}$$

and for quantum computing, the complexity, denoted $C_{\text{QISD}}$, is given by

$$C_{\text{QISD}} \sim \sqrt{\frac{\binom{N}{K}}{0.29\binom{N-t}{K}}} \tag{39}$$

where each quantum iteration involves a function evaluation requiring $O(N^3)$ qubit operations.

We now apply Equation (38) to evaluate the complexity of a Goppa-code variant, specifically the (4096, 3556, 45) code, and compare it with that of an MCC offering similar key lengths. Notably, MCC provides an additional layer of security: while the parameter $e$ is publicly known, the actual number of errors $t$ remains unknown. Furthermore, the decoder's use of polynomial division introduces additional, unpredictable errors, further obscuring the effective error count. Consequently, whereas ISD algorithms assume a fixed value of $t$, decoding in MCC entails iterating over a range of possible error weights. For a meaningful comparison, however, we assume that in the MCC scheme, the effective number of errors is fixed at $eN + \alpha$, where $\alpha$ represents the additional errors introduced during polynomial division, as defined in Equation (32). Substituting the parameters of the Goppa (4096, 3556, 45) code we obtain:

$$C_{\text{ISD}}(\text{Goppa}) \approx \frac{\binom{4096}{3556}}{0.29\binom{4096-45}{3556}} \approx 7.06\times10^{40} \tag{40}$$

For an MCC with a comparable key length, we can choose parameters $N$ =5600 and $K$= 2400. For instance, by selecting polynomials $\boldsymbol{p}_0(x)$ and $\boldsymbol{p}_1(x)$of degree 14 and, $\boldsymbol{q}_0(x)$ and $\boldsymbol{q}_1(x)$ of degree 386 (resulting in a rate ½ MCC with memory length 400 and 2400 input bits) the corresponding 5600-bit codeword achieves a public key length comparable to Goppa code, approximately 2 megabytes in both cases. For these parameters, selecting $\boldsymbol{q}_0(x) = x^{193}$ and $\boldsymbol{q}_1(x) = 1+x^{386}$ with $e$ = 0.02 yield an error probability at the Viterbi decoder input of:

$$\frac{eN + \alpha}{N} \approx 0.0715$$

This corresponds to an average of approximately 400 errors in a 5600-bit ciphertext. Such an error rate lies well within the correction capability of a rate ½ CC with memory length 14, making decoding failures highly improbable. By contrast, incorrectly unmasked candidates are decoded (as analysed in Section 6) to codewords lying at a Hamming distance of approximately

$$0.11\times5600 \approx 616$$

from the received vector. This distance exceeds the expected correction radius by a wide margin, placing such candidates outside the decoder's error sphere and thereby making them easily distinguishable from the correct solution. The separation implied by the code's distance spectrum ensures that the valid plaintext can be identified with overwhelming probability, while incorrect candidates are systematically rejected. Under these assumptions, the decoding complexity of the MCC is:

$$C_{\mathrm{ISD}}(\mathrm{MCC}) \approx \frac{\binom{5600}{2400}}{0.29\binom{5600-400}{2400}} \approx 5.2\times 10^{102} \quad (41)$$

This result reflects a security improvement by a factor of approximately $7.35\times10^{61} \approx 0.7\times 2^{206}$ compared to the classic McEliece scheme in Equation (40). For quantum ISD (QISD), the improvement is roughly $0.35\times 2^{103}$, consistent with the square-root speedup provided by Grover's algorithm. Increasing the public key length further enhances resistance to cryptanalysis, as indicated by Equation (38). Given that the primary role of the MCC cryptosystem is to securely distribute a key for subsequent reuse, a key length of several megabytes is not considered a significant limitation, except in cases where the data message itself is very short.

To estimate the probability of a decoding failure (and thus the need for retransmission), consider the following example. Suppose we employ a rate ¼ CC with memory length 10, defined by:

$$G_P(x) = [2327, 2313, 2671, 3175] \quad (42)$$

where the polynomials are expressed in octal form. This code provides robust error correction capabilities, characterised by a free distance of $d_{free} = 29$. In practical terms, this means that the Hamming distance between any two paths through the 1024-state trellis—which diverge from a common state and remerge after 11 segments (corresponding to 44 bits)—is at least 29. Consequently, the CC can reliably correct up to 14 errors within each 44-bit window. Next consider the following high-memory polynomials:

$$G_Q(x) = [1+x^{495}+x^{990}, x^{247}, x^{743}, 1+x^{990}].$$

For an error rate $e = 0.04$, simulations yield

$$\frac{\alpha}{N} \approx \frac{0.18+0+0+0.13}{4} \approx 0.0775$$

Thus

$$e + \frac{\alpha}{N} = 0.04 + 0.0775 = 0.1175$$

The Viterbi decoder fails if more than 14 errors occur within a 44-bit window. At an effective error rate of 0.1175, the probability of such an event is approximately $8.998\times10^{-5}$. Over 500 windows the probability of successful decoding is therefore $(1 - 8.998\times10^{-5})^{500} \approx 0.956$.

This result demonstrates that the MCC cryptosystem maintains a high decryption success rate even for ciphertexts spanning tens of thousands of bits.

We note that low-rate public codes can correct a larger fraction of errors than high-rate codes, but their higher redundancy can impact both the efficiency and security of code-based cryptosystems, as indicated by Equation (38). When $K \ll N$, the large number of parity constraints $N$-$K$ relative to the small number $K$ of message bits may be leveraged by an attacker to locate low-weight codewords or speed up decoding, as shown in [12]. Consequently, the code rate should be a key factor when choosing the underlying CC construction.

Finally, we examine the computational complexity of decryption. Schöffel *et al.* [13] present a hardware/software co-design implementation of a Hamming Quasi-Cyclic (HQC) cryptosystem tailored for IoT edge devices, providing a detailed evaluation of energy consumption, performance, and deployment trade-offs. The study demonstrates that code-based schemes can constitute practical alternatives to lattice-based approaches in resource-constrained environments. In [14], another hardware accelerator for HQC-based post-quantum cryptography is introduced, implementing key generation, encapsulation, and decapsulation on an FPGA with substantial performance gains achieved through tight integration with a RISC-V core.

In the proposed MCC cryptosystem, the primary contributor to decryption complexity is the Viterbi decoding stage. Other operations—such as polynomial division, unmasking, interleaving, and deinterleaving—are applied to the entire received ciphertext but incur a negligible computational cost compared to the per-bit processing requirements of the Viterbi algorithm. To quantify this cost, we evaluate the number of Add-Compare-Select (ACS) modules utilized per decrypted plaintext bit. These modules can be efficiently realized in both software and hardware, enabling practical and scalable deployment.

Suppose the MCC employs a CC with memory $p$ and a set $\mathcal{L}$ consisting of $l$ random masking vectors. In this case the decryption process requires $2^l$ parallel Viterbi decoders, each utilizing $2^p$ ACS modules per bit. Consequently, the total number of ACS modules per plaintext bit is $2^{l+p}$. This complexity measure scales linearly with the plaintext length $K$, ensuring predictable and manageable performance as the data size grows. For example, using the CC defined in Equation (42) with $l = 5$ and memory $p = 10$, the total ACS operations per bit become $2^{l+p} = 2^{15} = 32{,}768$, a computational load well within the capabilities of commercial processors, including those found in modern mobile devices.

## 9. CONCLUSION

This work introduces a novel post-quantum encryption scheme based on high-memory masked convolutional codes, overcoming key limitations of traditional block code-based approaches. Security is reinforced through semi-invertible transformations that produce fully dense, random-like generator matrices, mitigating vulnerabilities associated with low-weight or structured codes.

The scheme employs a dual-layer error injection mechanism: (i) high-rate deliberate random errors and (ii) additional noise inherently introduced through polynomial division. This layered strategy significantly increases resistance to cryptanalysis, even under conservative assumption that the adversary possesses full knowledge of the underlying convolutional code. The interplay of semi-invertible masking, polynomial division and invertible scrambling ensure that decryption without the private key remains computationally infeasible. Notably, the scheme achieves performance improvements over Classic McEliece by factors exceeding $2^{100}$ against quantum adversaries and $2^{200}$ against classical adversaries.

Beyond its strong security guarantees, the scheme supports arbitrary plaintext lengths, scales efficiently, and enables high-throughput implementation via parallel Viterbi decoders. These properties make the proposed construction a compelling candidate for both classical and post-quantum cryptographic deployments.

## AUTHOR

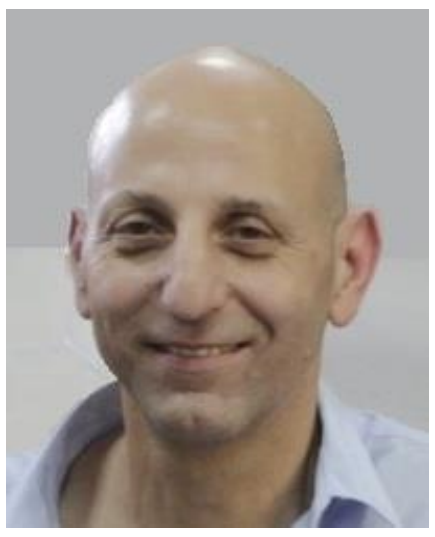

**Meir Ariel** received his B.Sc. and M.Sc. degrees (with honours) in Electrical Engineering and a Ph.D. in Algebraic Group Theory, all from Tel Aviv University. He has over 30 years of R&D experience in signal processing, wireless communications, and space technologies, with leadership roles across industry and the public sector. From 1999 to 2013, he was CEO of technology ventures in vision systems, VoIP, and fintech. Since 2013, he has served as Director General of the Herzliya Science Center, and in 2018, he founded Tel Aviv University's Space Engineering Center, leading 23 nanosatellite developments for space research and experimental communications. He holds 15 patents in information theory and signal processing and has served on national advisory boards in science and education. His honours include the Marco Polo Society Award (2018) and recognition by Israel's Ministry of Science and Technology as one of the country's 60 pioneering inventors (2016).